%% file: paper.tex
\documentclass[9pt]{IEEEtran}
\usepackage{graphicx}
\usepackage{tikz}
\usepackage{amsmath}
\usepackage{amssymb}
\usepackage{times}
\usepackage{subfigure}
\begin{document}

\sloppy

\pagestyle{plain}

\title{Sphinx: A Secure Architecture Based on Binary Code Diversification and Execution Obfuscation}
\author{Michel A. Kinsy, Donato Kava, Alan Ehret, Miguel Mark\\
Adaptive and Secure Computing Systems Laboratory \\
Boston University, Boston, USA
\vspace{-0.4in}}

\maketitle
\begin{abstract}
Sphinx, a hardware-software co-design architecture for binary code and runtime obfuscation. The Sphinx architecture uses
    binary code diversification and self-reconfigurable processing elements to maintain application functionality while
    obfuscating the binary code and architecture states to attackers. This approach dramatically reduces an attacker's
    ability to exploit information gained from one deployment to attack another deployment. Our results show that the
    Sphinx is able to decouple the program's execution time, power and memory and I/O activities from its functionality. It is also practical in the sense that the system (both software and hardware) overheads are minimal. 
\vspace{-0.2in}
\end{abstract}

% sections 
\input{intro}
\input{arch}
\input{eval}

\bibliographystyle{IEEEtran}
\bibliography{paper}

%\appendix
%\input{appendix}

\end{document}

%% file: intro.tex
\section{Introduction}
\label{sec:intro}

Over the last several decades, computers have become more and more important to society and our daily lives. Almost all modern activity, from banking to research to small talk, has been connected to computers and the Internet. Considering how much computers have permeated our lives, it is necessary that computer systems are secured against malicious activity. This need for secure computing systems has produced extensive research efforts in many areas of computer science and engineering.
Despite these efforts, the frequent reports of security breaches and cyber-attacks demonstrate that the work is on-going. One threat that has plagued security researchers and software developers for years is 
the possibility of reverse engineering programs and binary code to learn system vulnerabilities and modify the code to circumvent other security measures.  

Few hardware systems have been developed to protect software from reverse engineering. Those methods that have been proposed are often centered around encrypting the program and then decrypting before execution.  Depending on when the program is decrypted, an attacker may be able to use probes to obtain the plaintext program, such as from the bus between the CPU and the memory \cite{Kuhn1998}. One suggestion to avoid this attack is to decrypt the instructions immediately prior to execution as is done in \cite{Cappaert2008}. While this approach is promising in terms of blocking reverse engineering, it does not take into consideration side-channel attacks which could pose a threat.

Side Channel Attacks (SCAs) are a category of hardware attacks in which the attacker uses unintentional outputs (called side channels) to discover hidden information. Examples of side channel attacks include analyzing power, timing, and electromagnetic radiation to correlate the values with the secret information \cite{Zhou2005}. There are many proven examples of these attacks in \cite{barel2006sorcerer, hess2000information, fan2010state}. Side channel attacks provide a major attack surface because they observe unintentional output sources which are usually ignored by system and software designers. In order to fully hide a program from attackers, all side channels would have to be completely obscured.

Many counter measures have been suggested to help defend against side channel attacks. Some works present changes to algorithms that attempt to tax the processor similarly in all cases, obscuring the power, electromagnetic (EM), and time channels. Other suggestions include oblivious RAM (ORAM), which is a hardware structure that randomizes memory access to prevent eavesdroppers from discovering patterns in the memory access \cite{stefanov2011}. Secure processors, like ASCEND \cite{Fletcher2012} and AEGIS \cite{suh2014}, attempt to implement impenetrable security measures in hardware. ASCEND tries to block side-channel attacks by activating every hardware module on each clock cycle and accessing memory at regular intervals even if no access is pending. Such security measures do not come cheap; all of these systems suffer high cost and significant slowdown to execution.

Numerous security mechanisms have been proposed to prevent reverse engineering and to secure hardware against side channel attacks; however, each of the existing systems is found to be lacking in some way. The Sphinx system has a secure processor that can provide security against reverse engineering and side channel attacks without incurring excessive time and power overheads.

\begin{figure}[h]
\vspace{-0.1in}
\begin{center} 
\includegraphics[width=0.40\textwidth]{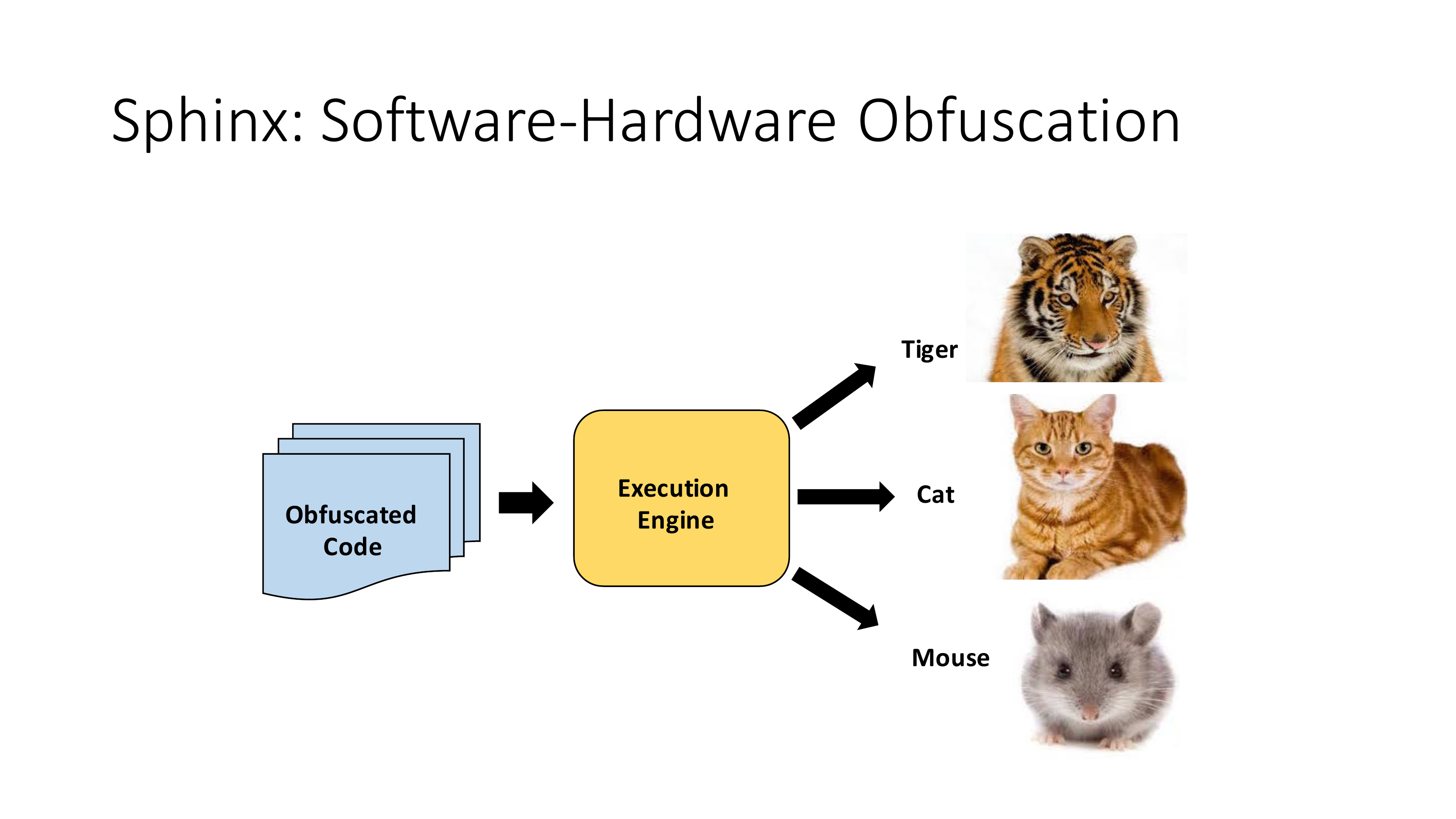}
\vspace{-0.1in}
\caption{Illustration of the Sphinx Architecture: The observed execution behavior depends on the context.}
\label{fig:illustration}
\end{center} 
\vspace{-0.3in}
\end{figure}

The general idea behind the Sphinx architecture is the minimization of attack-surface via software-hardware obfuscation. Every time a program (e.g., written in C or C++) is compiled, a different binary 
is generated. Every time a binary is executed, the power, execution time and memory activity profiles are all different. In essence, for the same application, the binary code and architecture states are obfuscated and the compute system operates differently to attackers. Figure \ref{fig:illustration} 
illustrates this behavior changing nature of the architecture. 

%% file: arch.tex
\section{ Obfuscation-based Secure Architecture Design}
\label{sec:arch}

\begin{figure}[h]
\includegraphics[width=0.45\textwidth]{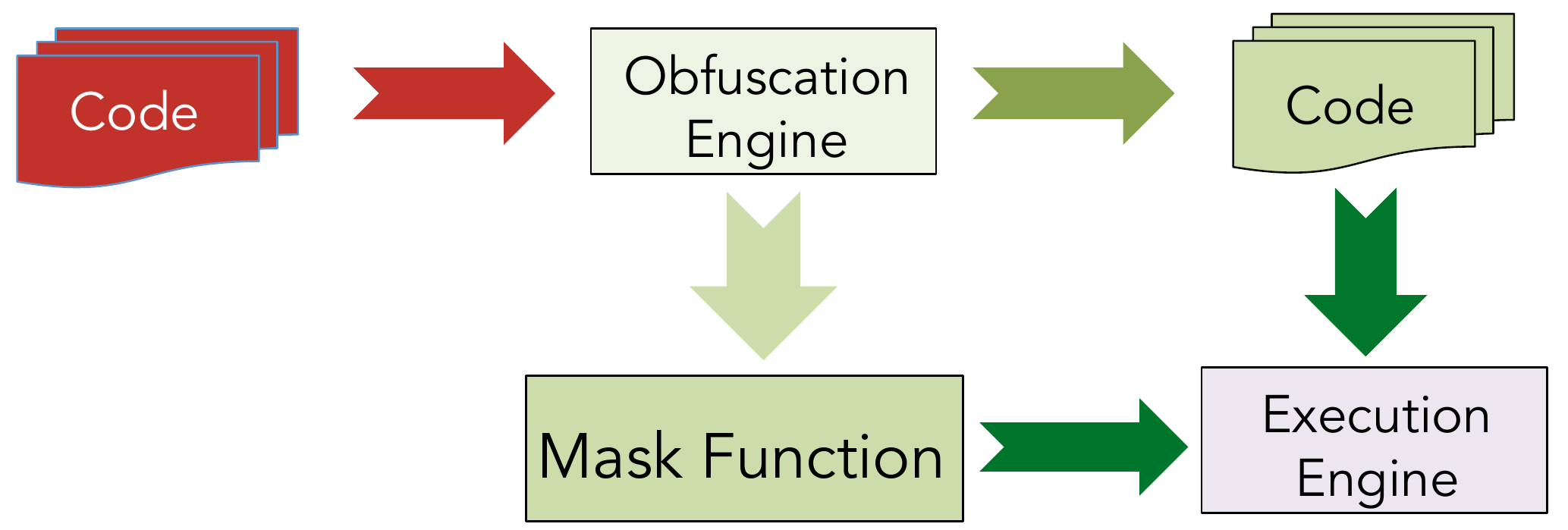}
\vspace{-0.1in}
\caption{Sphinx system functionality overview.}
\label{fig:function}
\vspace{-0.1in}
\end{figure}

\begin{figure}[!h]
\begin{center} 
\includegraphics[width=0.5\textwidth]{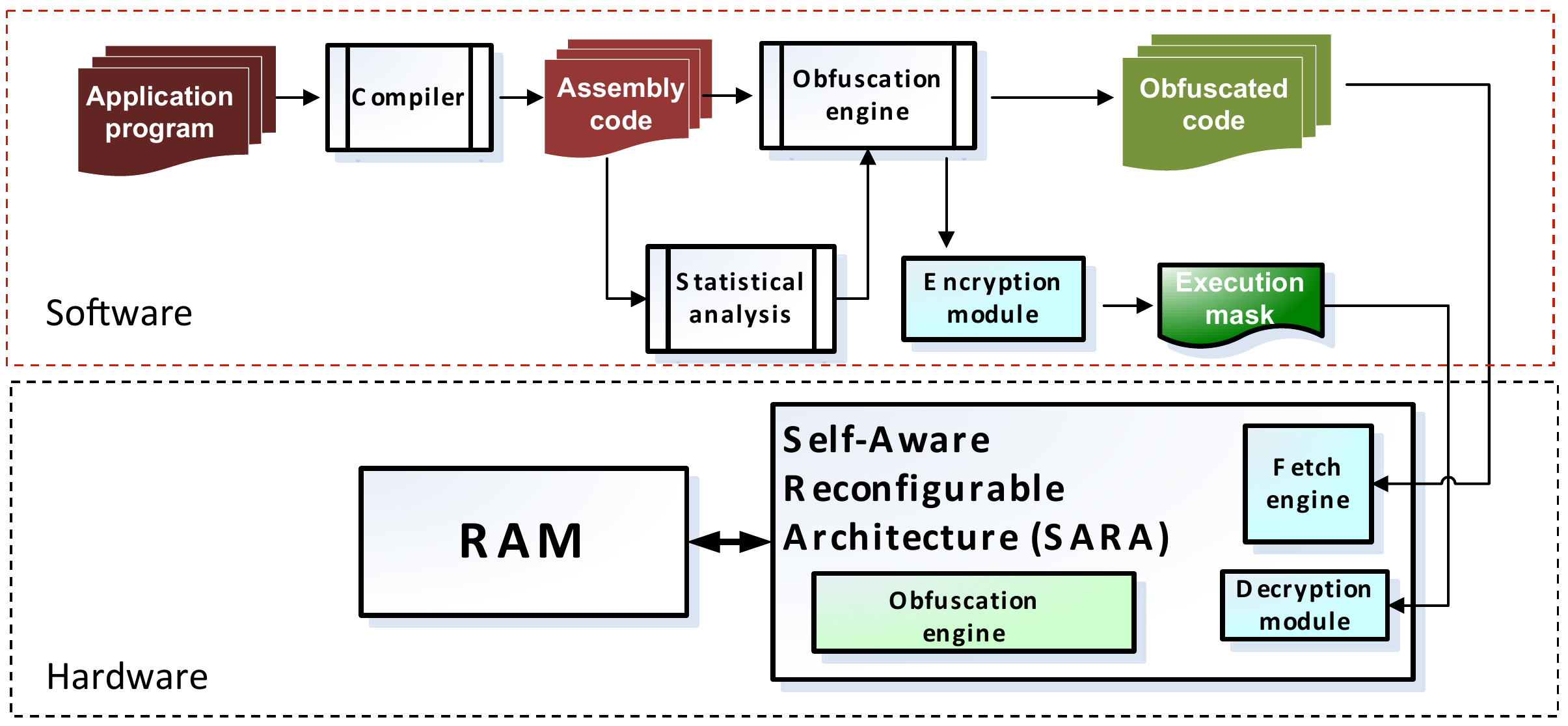}
\vspace{-0.1in}
\caption{Sphinx obfuscation-based secure system}
\label{fig:sphinx}
\end{center} 
\vspace{-0.1in}
\end{figure}

\begin{figure}[h]
\includegraphics[width=0.45\textwidth]{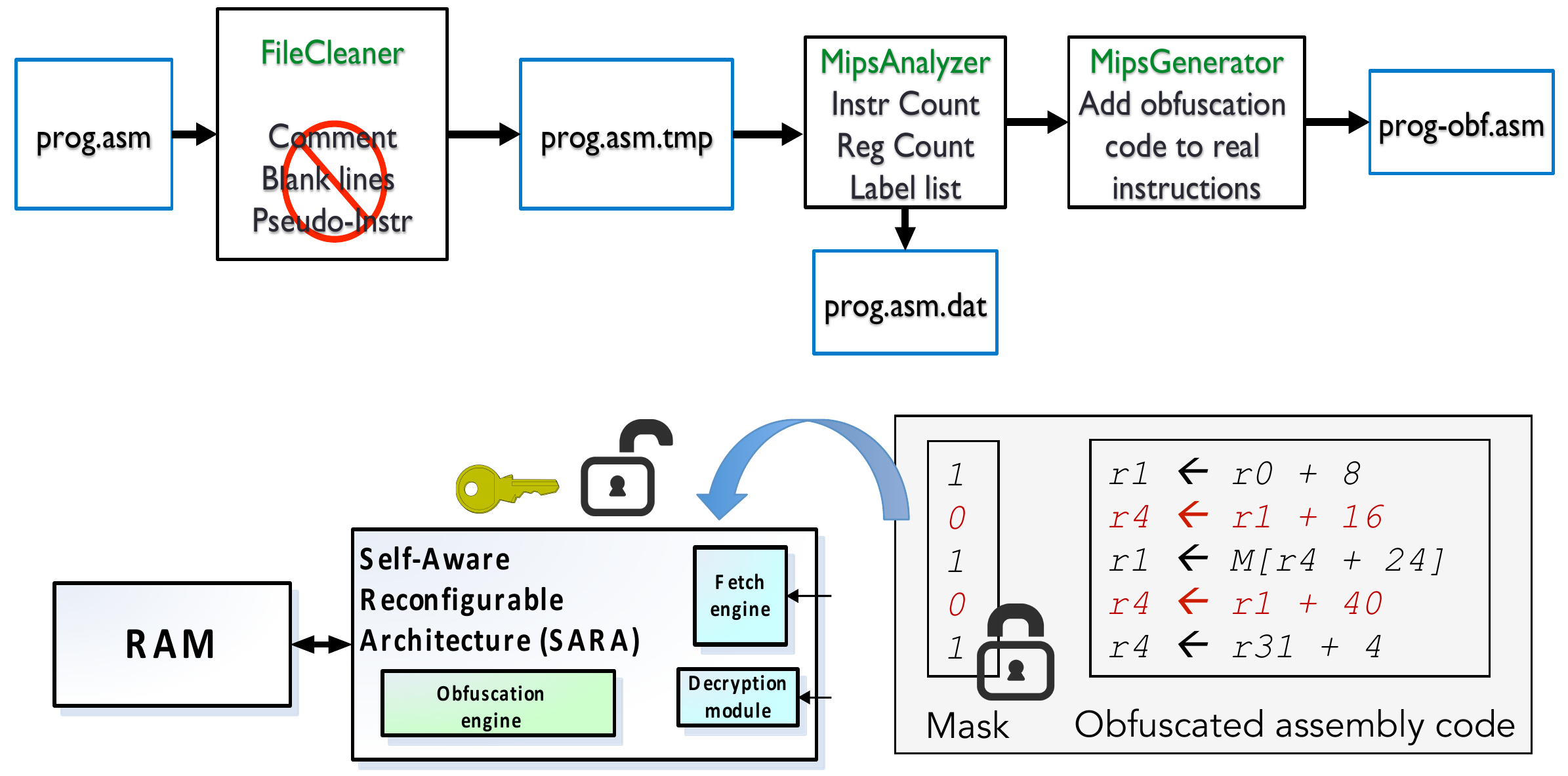}
\vspace{-0.1in}
\caption{The Sphinx software obfuscation process}
\label{fig:masking}
\vspace{-0.1in}
\end{figure}

Sphinx is a software-hardware co-design framework to efficiently execute a program while providing moving target defense against code reverse engineering and side-channel based attacks. Figure \ref{fig:function} depicts the system functionality. The Sphinx secure system has two parts: a software obfuscator and a hardware/processing unit. Figure \ref{fig:sphinx} shows the Sphinx system with its software and hardware components. 
The obfuscator inserts random machine instructions in the program at compile-time and creates a binary mask that
indicates which instructions are real and which ones are falsified. The goal of using obfuscation is twofold: first, to
render assembly or binary code analysis attacks harder - not impossible, since Barak et al., \cite{barak2010possibility}
have already shown that the impossibility of indistinguishable obfuscation; second, to give runtime adaptation range to the hardware. On the software side, the code is compiled and analyzed. The result of the analysis is used to generate a similar profile obfuscation assembly code to be added to the original. Figure \ref{fig:masking} shows the obfuscation process. The \textit{obf.asm} file goes through the normal compilation process. 
The binary mask is encrypted and loaded into the processor's memory with the obfuscated program. To run the program the processor transfers the encrypted mask to the execution unit and decrypts it using a securely stored key. In this work, we assume the secrecy of the key is provided by a Physical(ly) Unclonable Functions (PUF) technique. Then, the execution unit runs the program throwing out fake instructions as indicated by the mask. 

\begin{figure}[h]
\vspace{-0.1in}
\includegraphics[width=0.45\textwidth]{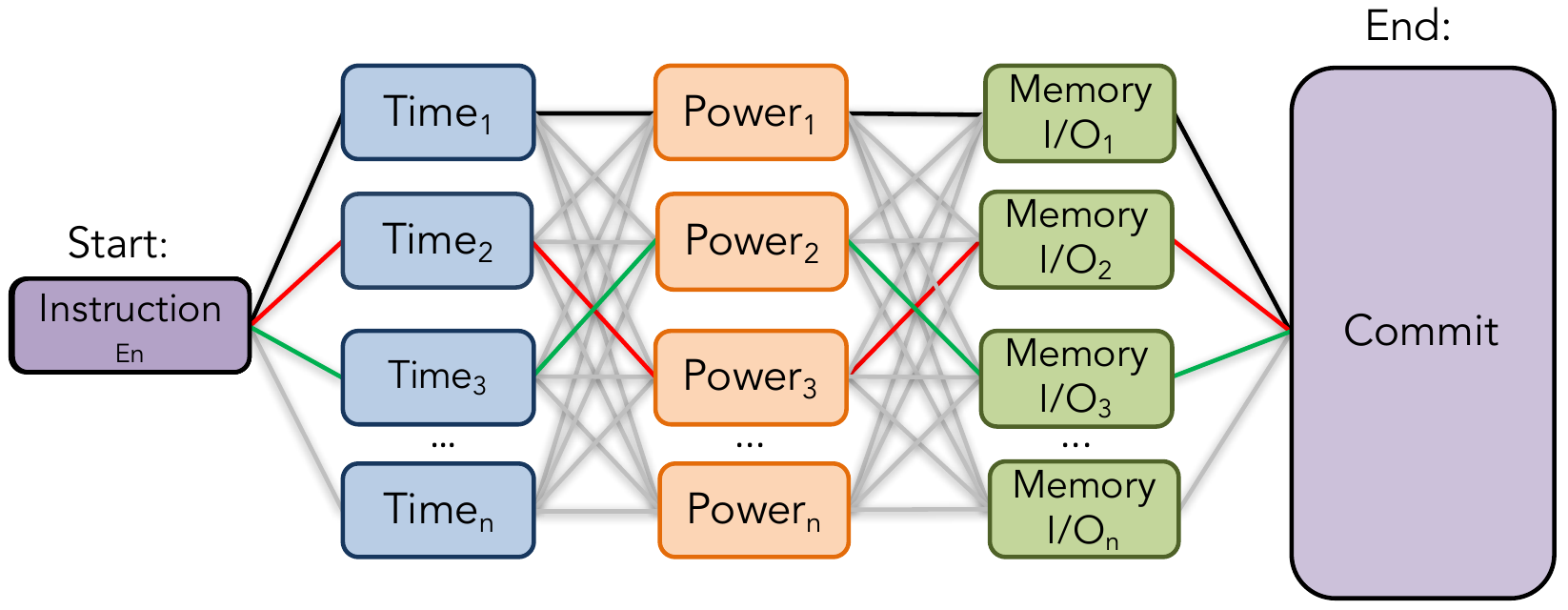}
\vspace{-0.1in}
\caption{Instruction execution profile candidates}
\label{fig:profiles}
\vspace{-0.2in}
\end{figure}

The Sphinx core execution unit is a Self-Aware Reconfigurable Architecture (SARA). The key idea for the SARA design approach is to allow the hardware to have multiple ways of executing the same instruction with different time, power and memory/IO profiles, illustrated in Figure \ref{fig:profiles}. The software level obfuscation gives reconfiguration range to the hardware and aids the architecture in isolating the program's execution time, power and memory and I/O activities from its functionality.  Figure \ref{fig:profiles} shows the effects of user-defined entropy levels on the distribution of real and falsified instructions. Effectively, the Sphinx architecture provides the following capabilities (a) performance/timing-awareness for timing obfuscation, (b) power-awareness for power obfuscation and (c) self-organized data storage for memory and I/O obfuscation. 

\begin{figure}[h]
\includegraphics[width=0.5\textwidth]{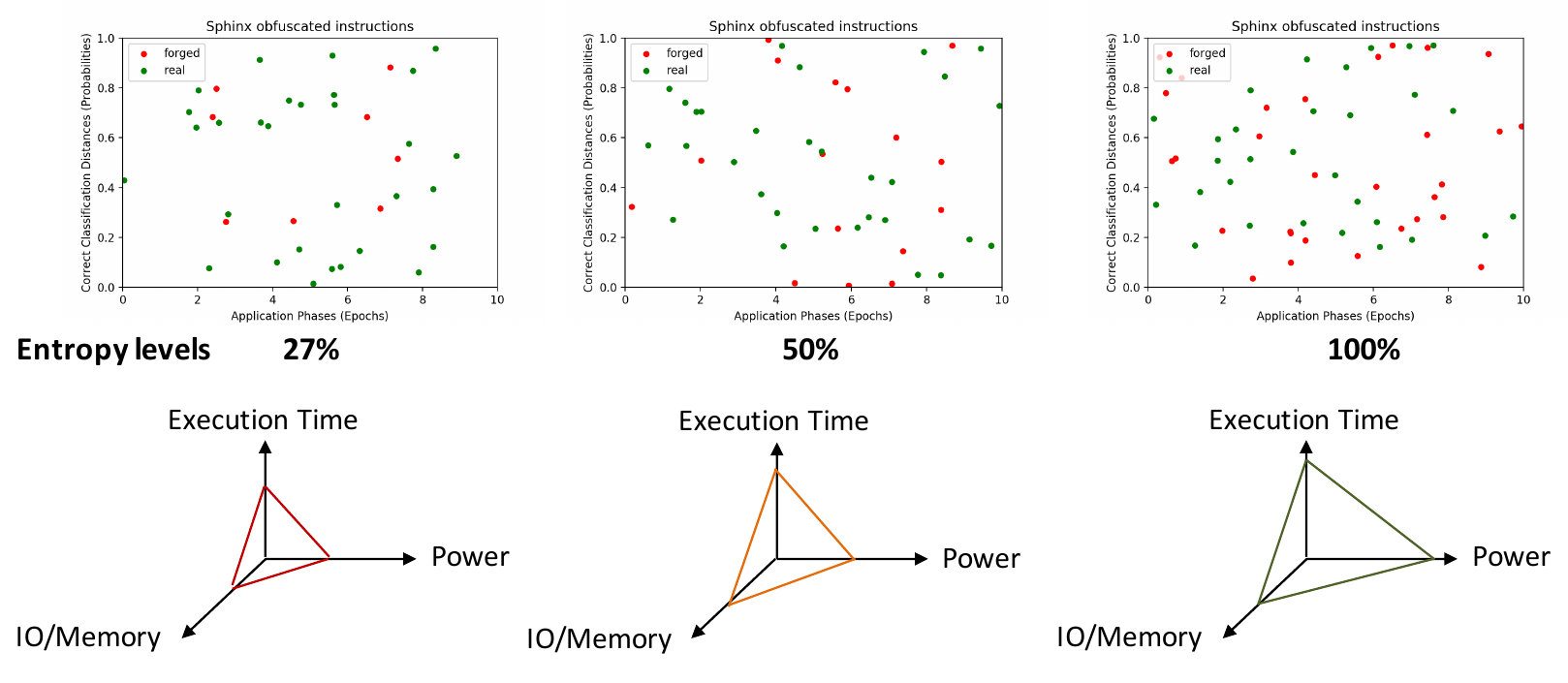}
\vspace{-0.2in}
\caption{Effects of user-defined entropy levels on the distribution of real and falsified instructions}
\label{fig:entropy}
\vspace{-0.1in}
\end{figure}

%% file: eval.tex
\section{Evaluations}
\label{sec:eval}
\begin{figure}[h]
\includegraphics[width=0.5\textwidth]{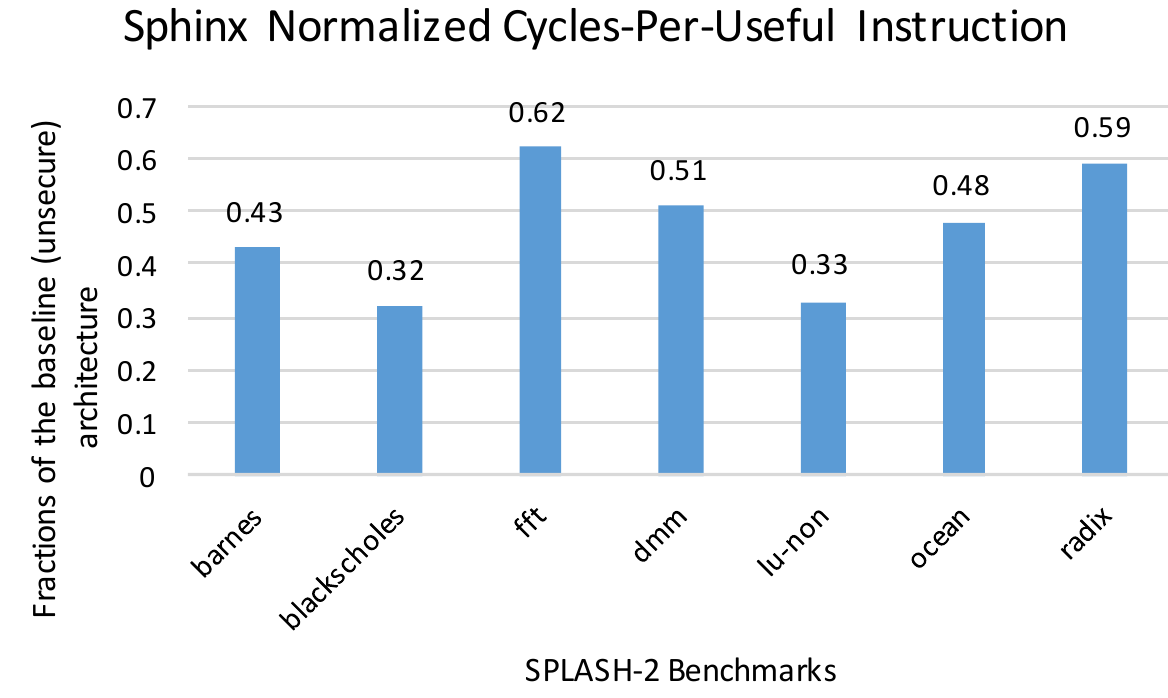}
\vspace{-0.2in}
\caption{Splash-2 benchmarks on the Sphinx system.}
\label{fig:results}
\vspace{-0.1in}
\end{figure}

To test the Sphinx system, we implement a Verilog version of the hardware. We use a number of common benchmark suites (SPLASH-2, PARSEC, SPEC CINT2006). The selected benchmark applications are compiled, obfuscated and executed on the emulated hardware. The benchmarks are also run on an unsecured but otherwise similar processor to perform a comparative study. Figure \ref{fig:results} highlights some of the performance results of the Sphinx system. 

